\documentclass{article}

\usepackage{microtype}
\usepackage{graphicx}
\usepackage{subfigure}
\usepackage{booktabs} 

\usepackage{hyperref}

\usepackage[accepted]{icml2023}

\usepackage{amsmath}
\usepackage{amssymb}
\usepackage{mathtools}
\usepackage{amsthm}

\usepackage[capitalize,noabbrev]{cleveref}

\theoremstyle{plain}

\theoremstyle{definition}

\theoremstyle{remark}

\usepackage[textsize=tiny]{todonotes}

\icmltitlerunning{Domain Adaptation via MME for Real/Bogus Classification of Astronomical Alerts}

\begin{document}

\twocolumn[
\icmltitle{Domain Adaptation via Minimax Entropy for Real/Bogus Classification of Astronomical Alerts}



\icmlsetsymbol{equal}{*}

\begin{icmlauthorlist}
\icmlauthor{Guillermo Cabrera-Vives}{DIICC,UDS,MAS}
\icmlauthor{César Bolivar}{DIICC}
\icmlauthor{Francisco Förster}{IDIA,MAS,CMM}
\icmlauthor{Alejandra M. Muñoz Arancibia}{MAS,CMM}
\icmlauthor{Manuel Pérez-Carrasco}{UDS,DIICC,MAS}
\icmlauthor{Esteban Reyes}{Fintual}
\end{icmlauthorlist}

\icmlaffiliation{DIICC}{Department of Computer Science, Universidad de Concepción, Concepción, Chile}
\icmlaffiliation{UDS}{Data Science Unit, Universidad de Concepción, Concepción, Chile}
\icmlaffiliation{MAS}{Millennium Institute of Astrophysics, Chile}
\icmlaffiliation{IDIA}{Data and Artificial Intelligence Initiative (IDIA), Faculty of Physical and Mathematical Sciences, University of Chile, Chile}
\icmlaffiliation{CMM}{Center for Mathematical Modeling (CMM), Universidad de Chile, Chile}
\icmlaffiliation{Fintual}{Fintual Administradora General de Fondos S.A., Santiago, Chile}

\icmlcorrespondingauthor{Guillermo Cabrera-Vives}{guillecabrera@inf.udec.cl}

\icmlkeywords{Machine Learning, ICML}

\vskip 0.3in
]



\printAffiliationsAndNotice{}  

\begin{abstract}
Time domain astronomy is advancing towards the analysis of multiple massive datasets in real time, prompting the development of multi-stream machine learning models. 
In this work, we study Domain Adaptation (DA) for real/bogus classification of astronomical alerts using four different datasets: HiTS, DES, ATLAS, and ZTF. We study the domain shift between these  datasets, and improve a naive deep learning classification model by using a fine tuning approach and semi-supervised deep DA via Minimax Entropy (MME). We compare the balanced accuracy of these models for different source-target scenarios. 
We find that both the fine tuning and MME models improve significantly the base model with as few as one labeled item per class coming from the target dataset, but that the MME does not compromise its performance on the source dataset. 
\end{abstract}

\section{Introduction}
\label{sec:intro}

Time-domain survey telescopes are providing astronomers with vast amounts of data on celestial objects and phenomena. Surveys such as the Asteroid Terrestrial-impact Last Alert System \citep[ATLAS; ][]{tonry2018atlas} or the Zwicky Transient Facility \citep[ZTF; ][]{bellm2018zwicky} emit when the brightness or location of a source change, producing a continuous astronomical alert stream. The aggregation, annotation, and classification of alerts in a rapid and consistent
fashion is done by astronomical alert brokers \citep{narayan2018machine, nordin2019transient, smith2019lasair, forster2021automatic, moller2021fink}. An important number of these alerts are \emph{bogus} artifacts created by the image reduction pipelines, hence, the importance of creating real/bogus classification algorithms which have proven to be extremely useful for detecting real astrophysical phenomena. During the last decade, most of these algorithms have been based on Convolutional Neural Networks \citep{cabrera2016supernovae, cabrera2017deep, reyes2018enhanced, duev2019real, turpin2020vetting, yin2021supernovae, rabeendran2021two} which need a significant amount of data to be trained. Domain adaptation (DA) techniques such as the Minimax Entropy \citep[MME; ][]{saito2019semi} approach are an alternative that help training such models with fewer amount of data. Furthermore, DA allows models to perform inference simultaneously for multiple dataset that may follow different distributions. This is particularly important when developing multi-stream models for alert streams from next-generation telescopes such as the Vera Rubin Observatory as soon they start producing data. By effectively working across various alert streams and accounting for domain shifts, these models can leverage labeled and unlabeled data from multiple domains, enhancing their learning capabilities. Moreover, conducting inference on multiple alert streams using a single model facilitates performance monitoring across surveys.

In this work, we evaluate the use of fine tuning and MME for the real/bogus classification of alert stamps and their availability to transfer knowledge from models trained on \emph{source} surveys to different \emph{target} surveys using few shots of labeled sources. We start by describing the four datasets we use (HiTS, DES, ATLAS, and ZTF) in Section \ref{sec:data}. In Section \ref{sec:model}, we provide a comprehensive description of our feature extraction and classification models, as well as the domain adaptation techniques employed. We outline the details of our experiments in Section \ref{sec:experiments}, followed by the presentation of the obtained results in Section \ref{sec:results}. Finally, we draw conclusions based on these findings in Section \ref{sec:conclusions}.

\section{Data}
\label{sec:data}
We use image stamps from four surveys: the High Cadence Transient Survey \citep[HiTS; ][]{hits}, the Dark Energy Survey \citep[DES; ][]{goldstein}, ATLAS \citep{tonry2018atlas}, and ZTF \citep{ztf}. These four datasets consist of astronomical alerts represented as 3-channel images: 1) a reference image, 2) a science image taken at the time of observation, and 3) a difference image created by matching the point-spread-function of the science and reference images and subtracting them. 
Each alert within every dataset was assigned a corresponding label, indicating whether it is deemed "real" (representing an astronomical event of interest) or "bogus" (indicating a false detection). All images were cropped to $21\times 21$ pixels and were normalized to have a mean of 0 and a standard deviation of 1.

The primary goal of the HiTS survey was to detect supernovae during their earliest hours of explosions. Their real/bogus dataset consist of a total of 1,437,684 images of 21 $\times$ 21 pixels \citep{cabrera2017deep}. Bogus stamps were directly taken by the Dark Energy Camera \cite{flaugher2015dark} while real stamps were simulated within their pipeline. By construction, this dataset contains a total of 718,842 ``real'' stamps and 718,842 ``bogus'' stamps.

The DES dataset was obtained from \citealt{goldstein}\footnote{\url{https://portal.nersc.gov/project/dessn/autoscan/}} and it contains 51$\times$ 51 pixels stamps from 898,963 source candidates. Of these candidates, 454,092 are simulated supernovae labeled as "real", and 444,871 are "bogus" sources that came out of the DES pipeline 
\citep{abbott2018dark}.

ATLAS is a sky survey system that aims at finding dangerous near-Earth asteroids. We use 61$\times$61 pixels stamps coming from 3,678 candidate sources. This dataset was visually labeled and is composed of 500 persistent burn trails, 500 cosmic rays, 500 spike artifacts, 500 noise fluctuations, 500 sources labeled as asteroids, and 678 candidates labeled as asteroid streaks. The "real" dataset was created by joining the labeled asteroids and asteroid streaks, while the "bogus" class was created by combining the burn trails, cosmic rays, spikes, and noise.


We gathered ZTF stamps following the procedure described by \citealt{carrasco2021alert}. The raw dataset consists of 63 $\times$ 63 pixels stamps for a total of 36,262 source candidates, but 467 images were of a smaller resolution and were discarded. This dataset originally had 9,996 images labeled as active galactic nuclei, 1,079 labeled as supernovae, 9,938 labeled as variable stars, 9,899 labeled as asteroids and 5,350 labeled as bogus. All the non bogus labels were combined into the single ``real'' label. 
Some images contained bad pixels, that were  replaced by the median of the image where they were found.


\section{Model}
\label{sec:model}
Our baseline classification model was taken from \citealt{dann} and consists of a \textit{feature extractor} component and a \textit{predictor} component. The feature extractor component is composed of two 2-dimensional convolutional layers, each followed by a max-pooling layer. Batch normalization and ReLU activation functions are applied to both convolutional layers. The predictor component consists of three linear layers, each using batch normalization and ReLU activation functions, with the exception of the last layer which employs a Softmax function. As a benchmark, this model was trained with each dataset separately using a binary cross entropy loss function.

The MME model, similar to the baseline model, comprises a feature extractor and a predictor component. The feature extractor component shares the same architecture as the base model. The predictor component includes a L2 normalization layer, succeeded by a linear layer scaled by a temperature hyperparameter, and a Softmax activation function. Each class is represented in the feature space as an estimated vector \emph{prototype}. This model is trained using two datasets: a fully labeled \textit{source} dataset and a partially labeled \textit{target} dataset. It is worth mentioning that while MME allows for unsupervised learning using the target dataset, our work focuses on the semi-supervised approach. During training, the model parameters are optimized using a two-term loss function:
\begin{equation}
\mathcal{L} = H (y_s , \hat{y}_s ) + \lambda H ( \hat{y}_u ), \label{eq:MME_loss}
\end{equation}
where $H (y_s , \hat{y}_s )$ represents the cross-entropy loss between the true ($y_s$) and predicted ($\hat{y}_s$) labels for the labeled (supervised) dataset, and $H ( \hat{y}_u )$ denotes the entropy of the predicted labels for the unlabeled (unsupervised) dataset. The weight $\lambda$ controls the balance between the two terms in the loss function. A gradient reversal layer \cite{dann} is inserted between the feature extractor and predictor components of the model. This layer flips the sign of the gradient value during backpropagation, but only the unlabeled data passes through this layer. 
Consequently, the entropy term is minimized for the unlabeled target examples, encouraging the model to learn discriminative features that cluster around the estimated prototypes, while its maximization encourages feature representations that are invariant to domain shifts. This mechanism helps the model effectively adapt to the target dataset, leveraging information from both the labeled and unlabeled data sources.



\begin{table*}[h]
	\centering
	\begin{tabular}{ccccc}
		\toprule
		Source / Target & HiTS & DES & ATLAS & ZTF \\
		\midrule
		HiTS  & \textbf{0.983 ± 0.004} & 0.811 ± 0.026 & 0.626 ± 0.019 & 0.548 ± 0.015 \\
        DES & 0.945 ± 0.011 & \textbf{0.955 ± 0.003} & 0.703 ± 0.006 & 0.606 ± 0.011 \\
        ATLAS & 0.777 ± 0.031 & 0.697 ± 0.058 & \textbf{0.967 ± 0.008} & 0.502 ± 0.036\\
		ZTF & 0.765 ± 0.023 & 0.752 ± 0.022 & 0.633 ± 0.019 & \textbf{0.945 ± 0.007} \\
		\toprule
	\end{tabular}
	\caption[Baseline results]{Baseline results. Each row/column corresponds to the mean and standard deviation of the balanced accuracy for that source/target experiment, calculated across 10 different random splits.
 }
	\label{tab:baseline}
\end{table*}

\section{Experiments}
\label{sec:experiments}

Three sets of experiments were defined: a \textit{baseline} training, a \textit{fine tuning} training and a \textit{domain adaptation} training. A single round of these experiments use the same training/validation/testing set partitioning. To compare the models performance, the balanced accuracy metric (average recall) was used. In order to avoid the overuse of target labels, 10 labeled items per class were used for validation for the fine-tuning and MME experiments. To address imbalanced data, oversampling was applied to prevent bias towards the overrepresented class.

The baseline model was trained independently for each separate source dataset. We then used these models to evaluate their performances on all four datasets. 
The fine tuning training, consists of taking a baseline model and further training it using a smaller labeled set from the other three target datasets that were not trained on. We perform fine tuning using 1, 5, 10, 20 and 40 labeled items per class. We evaluate the performance of the fine-tuned models in both the source and target datasets in order to evaluate their domain adaptation capacities. 

For DA via MME, we used the full labeled source training set, a small number of labeled items per class sampled from the target dataset, and an unlabeled dataset comprising the remaining target objects. The feature extractor weights are initialized with those of the corresponding baseline model following the approach of \cite{saito2019semi}. Different instances of MME are run for each source-target scenario, varying the amount of target labeled data (1, 5, 10, 20, and 40 items per class, the same as the fine-tuning experiment). To assess the domain adaptation capabilities of MME, we evaluate its performance on both the source and target data. The optimal value for the hyperparameter $\lambda$ in Eq. \ref{eq:MME_loss} is determined by training MME with multiple values (0.01, 0.02, 0.03, 0.05, 0.1, 0.5, and 10) and selecting the one yielding the highest balanced accuracy on the target validation set. All aforementioned experiment instances were repeated in a 10-fold manner, with each iteration employing a distinct random data partitioning.

\section{Results}
\label{sec:results}

We start by evaluating the transferring learning capability of the baseline models when trained on each dataset separately by calculating their performances when applied on stamps from all surveys. Table \ref{tab:baseline} shows the balanced accuracy for these experiments. All models achieve an accuracy over 94\% over the source dataset. Our results are consistent with the literature for DES \citep[$\sim$96\%, ][]{acero2022there} and ATLAS \citep[$\sim$95.2\%, ][]{rabeendran2021two} while for the other datasets we achieve a slightly lower accuracy than the state-of-the-art ($\sim$99.5\% for HiTS, \citealt{cabrera2017deep}; $\sim$98\% for ZTF, \citealt{duev2019real}). We attribute this decrease in performance to the capacity of our model, yet we deem it of lesser significance in light of the primary focus of this paper, which is the evaluation of domain adaptation techniques. We notice that the greater transferring capacity of models is achieved by DES$\rightarrow$HiTS (DES as source, HiTS as target) and HiTS$\rightarrow$DES, which is to be expected given that both datasets were obtained using the Dark Energy Camera and both aimed at searching for supernovae. The rest of the experiments show worse performances, posing the need of transfering these models to the target datasets.

\begin{figure*}
    \begin{tabular}{ccc}
         \includegraphics[width=.333\textwidth]{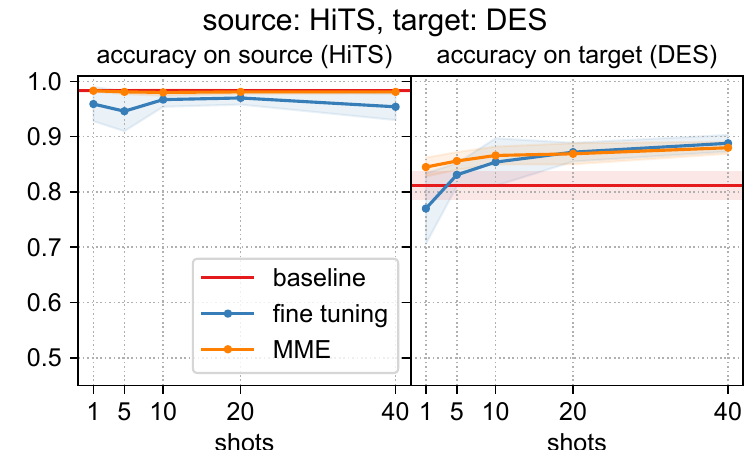} & \includegraphics[width=.333\textwidth]{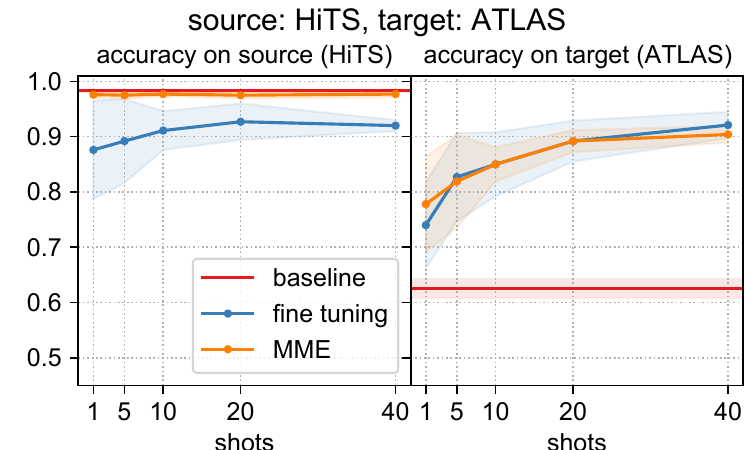} & \includegraphics[width=.333\textwidth]{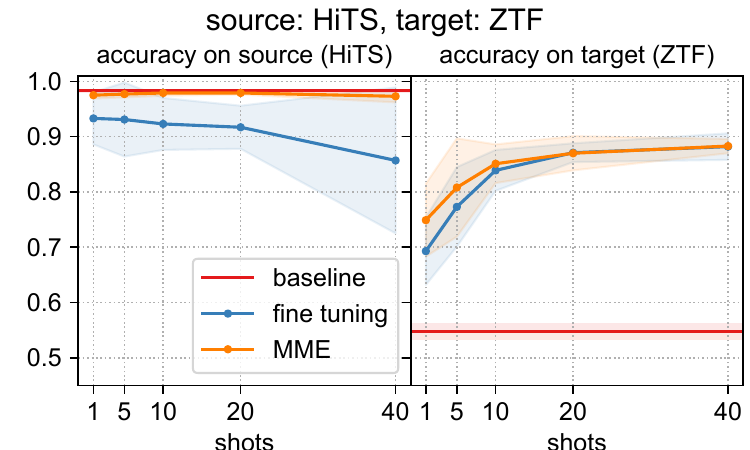} \\
         \includegraphics[width=.333\textwidth]{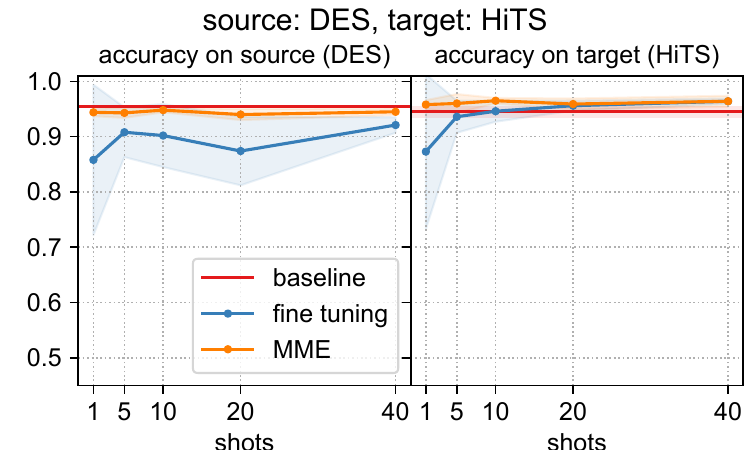} & 
         \includegraphics[width=.333\textwidth]{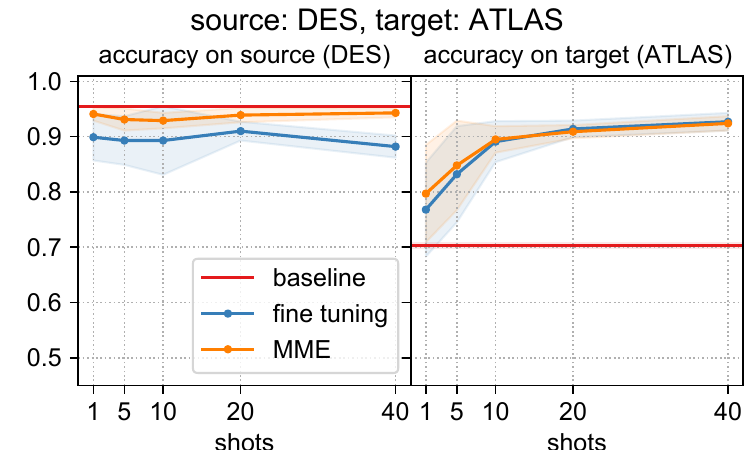} & 
         \includegraphics[width=.333\textwidth]{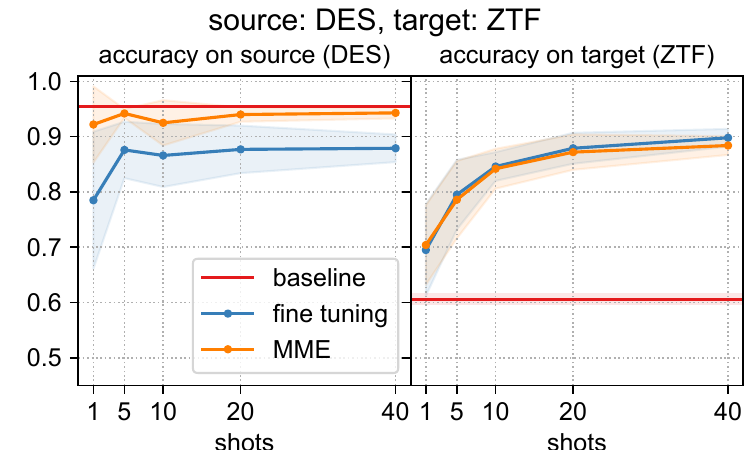} \\
         \includegraphics[width=.333\textwidth]{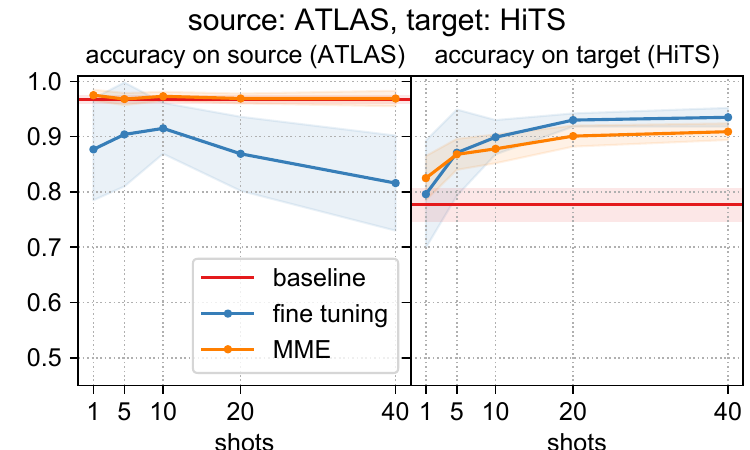} & 
         \includegraphics[width=.333\textwidth]{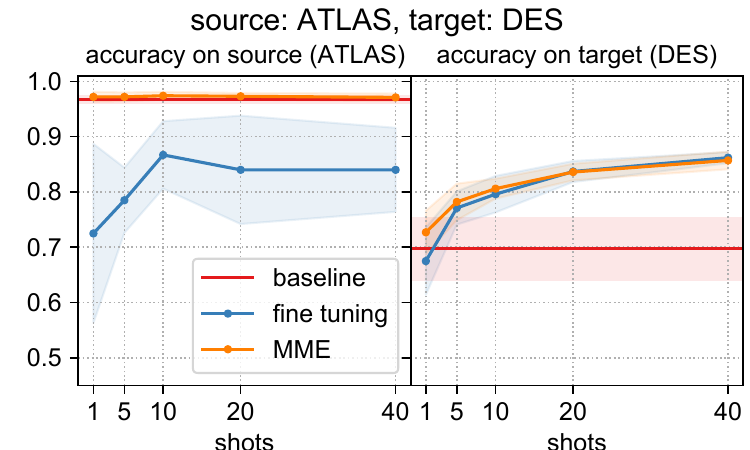} & 
         \includegraphics[width=.333\textwidth]{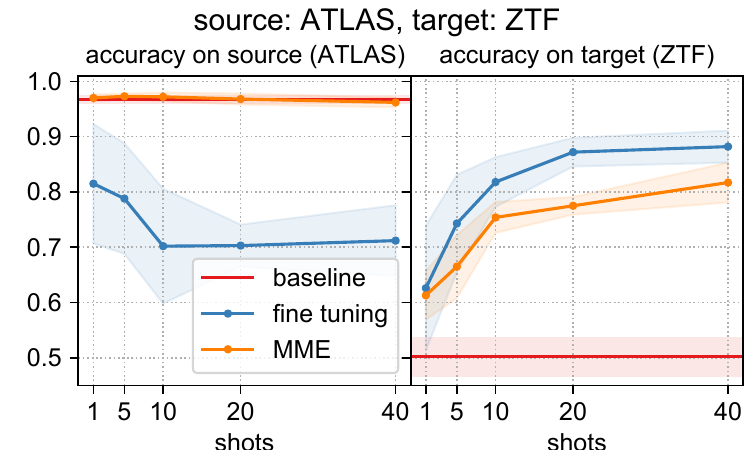} \\
         \includegraphics[width=.333\textwidth]{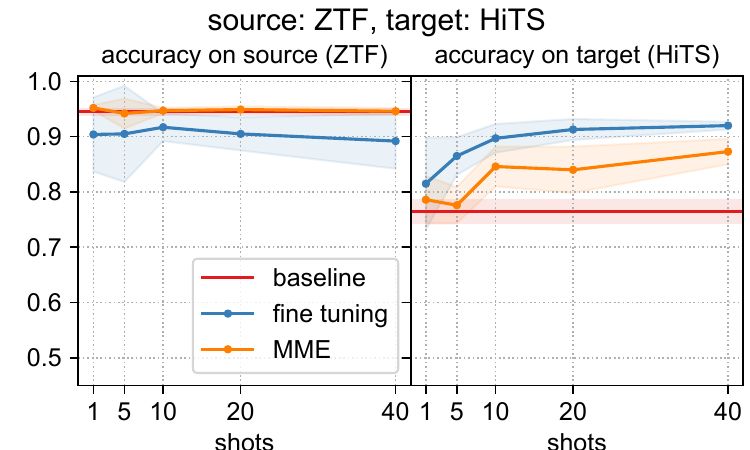} & 
         \includegraphics[width=.333\textwidth]{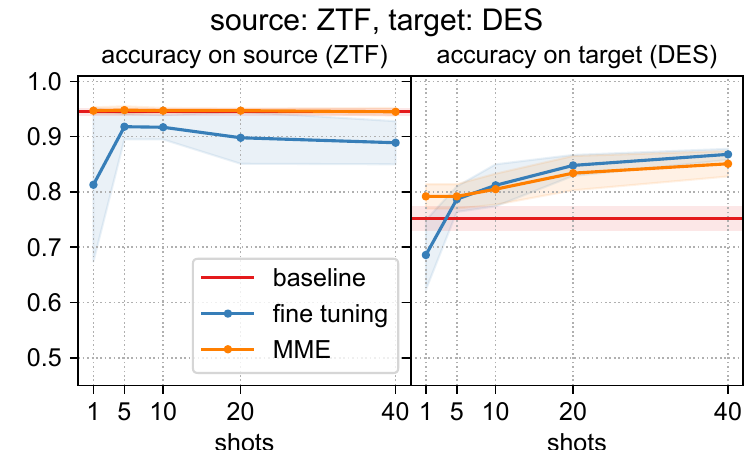} & 
         \includegraphics[width=.333\textwidth]{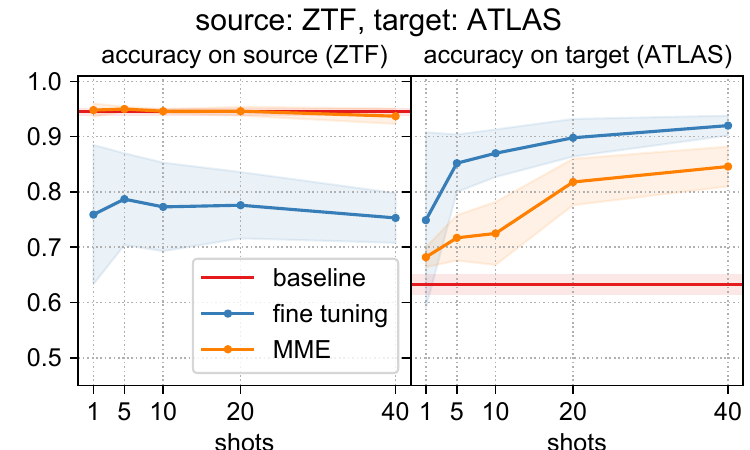} \\
    \end{tabular}
    \caption{Balanced accuracy for different source/target scenarios for HiTS, DES, ATLAS, and ZTF. Each pair of plots represent one source$\rightarrow$target experiment. The left plot shows the performance of fine tuning (blue) and MME (orange) on the source dataset as the number of shots increase, while the right plot shows their performance on the target dataset. The baseline result from Table \ref{tab:baseline} are shown as a horizontal red line.}
    \label{fig:DA}
\end{figure*}

We compare the capacity of transferring the learned representations to other datasets by using fine tuning and MME in Figure \ref{fig:DA} in terms of the number of labeled target objects presented to the model (shots). Each plot represents a source/target scenario. We show the accuracy of fine tuning and MME in terms of the shots on the source data (left plot of each panel) and on the target data (right plot of each panel). As noticed previously when discussing Table \ref{tab:baseline}, when transferring DES$\rightarrow$HiTS, fine tuning and MME are able to achieve competitive performances with few shots both for source and target stamps. In the target data, both fine tuning, and MME are able to surpass the baseline after only one or five shots of labeled objects from the target. Fine tuning and MME achieve comparable results on the target for most experiments, the exception being ATLAS$\rightarrow$ZTF, ZTF$\rightarrow$HiTS, and ZTF$\rightarrow$ATLAS where fine tuning outperforms MME. However, it is worth noting that fine-tuning the model on the target data consistently leads to a deterioration in performance for the source dataset, regardless of the number of shots employed. This is of particular importance when aiming at developing generalist models for multi-stream classification.

\section{Conclusions}
\label{sec:conclusions}

We evaluate the transferability of convolutional neural networks (CNN) trained to classify image stamps of astronomical alerts on a source dataset to target data coming from various surveys. Using real/bogus stamps coming from HiTS, DES, ATLAS, and ZTF, we show that even though the CNN models are able to achieve over $\sim$94\% balanced accuracy on the source dataset, they struggle to achieve competitive performances on stamps coming from surveys with a slightly different distribution. To address this, we examine two transfer learning techniques for the task: fine-tuning and domain adaptation via Minimax Entropy (MME). We show that both methods exhibit rapid learning capabilities from the target data with only a few labeled shots. However, MME maintains a high level of accuracy on the source domain, whereas fine tuning leads to a degradation of results on such domain. This is of special importance  when considering the training of generalist models capable of performing inference in a multi-stream scenario, especially in the context of the first-light of upcoming instruments like the Vera C. Rubin Observatory.

\section*{Acknowledgements}

The authors acknowledge support from the National Agency for Research and Development (ANID) grants: Millennium Science Initiative Program – ICN12 009 (GCV, FF, AMMA, MPC); FONDECYT regular 1231877 (GCV), 1200710 (FF); PIA/BASAL projects FB210005 (FF, AMMA), AFB-170001 (FF, AMMA), and FB210003 (FF).



\bibliography{MME_stamps}
\bibliographystyle{icml2023}

\newpage
\appendix
\onecolumn


\end{document}